\documentclass[11pt]{article} 
\usepackage[utf8]{inputenc}
\usepackage[T1]{fontenc}  
\usepackage{lmodern}      
\usepackage{geometry}
\usepackage{amsmath,amssymb}
\usepackage{booktabs}
\usepackage{enumitem}
\usepackage{hyperref}
\hypersetup{ colorlinks=true, linkcolor=black, citecolor=black, urlcolor=black, pdfborder={0 0 2}}
\usepackage{natbib} 
\newcommand{\textcite}[1]{\citet{#1}}
\usepackage{setspace}

\usepackage{pgfplots}
\pgfplotsset{compat=1.18}
\usepackage{xcolor}
\usepackage{tikz}
\usetikzlibrary{arrows.meta,positioning,fit,calc}
\definecolor{sdpBlue}{RGB}{40,90,150}     
\definecolor{sdpTeal}{RGB}{25,110,105}    
\definecolor{sdpOrange}{RGB}{190,120,40}  
\definecolor{sdpGray}{RGB}{95,95,95}      

\usepackage{authblk}

\title{Smart Data Portfolios:\\ A Governance Framework for AI Training Data}
\author[1,*]{A.\,Talha Yalta}
\author[2]{A.\,Yasemin Yalta}
\affil[1]{Department of Economics, TOBB University of Economics and Technology, Ankara, Turkiye}
\affil[2]{Department of Economics, Hacettepe University, Ankara, Turkiye}
\affil[*]{\textit{Corresponding author: yalta@etu.edu.tr}}
\date{\today}

\begin{document}
\onehalfspacing
\maketitle

\begin{abstract}
	Contemporary AI regulation, including the EU Artificial Intelligence Act and related governance frameworks, increasingly requires institutions to justify the training data used in automated decision-making. Yet existing governance regimes provide limited operational methods for selecting, weighting, and explaining data inputs. We introduce the Smart Data Portfolio (SDP) framework, which treats data categories as productive but risk-bearing assets, formalizing input governance as an information-risk trade-off. Within this framework, we define two portfolio-level quantities, Informational Return and Governance-Adjusted Risk, whose interaction characterizes attainable data mixtures and yields a Governance-Efficient Frontier. Regulators shape this frontier through risk caps, admissible categories, and weight bands that translate fairness, privacy, robustness, and provenance requirements into measurable constraints on data allocation while preserving model flexibility. A sectoral illustration shows how different AI services require distinct portfolios within a common governance structure. The framework provides an input-level explanation layer through which institutions can justify governed data use in large-scale AI deployment.
	\\ \textbf{Keywords:} AI governance, input governance, training data, portfolio theory, explainability, privacy, fairness
\end{abstract}

\section{Introduction}
\label{sec:introduction}

Artificial intelligence systems increasingly mediate access to credit, connectivity, health care, education, and public services. As automated decision-making expands, regulatory attention has shifted from model performance alone toward the governance of the data used to train and operate AI systems. Contemporary regulatory frameworks, including the European Union’s Artificial Intelligence Act \citep{EUAIAct2024} and the NIST Artificial Intelligence Risk Management Framework \citep{NIST2023}, now increasingly require institutions to document, assess, and justify the quality, representativeness, and provenance of training data. In practice, however, institutions confronting these obligations lack systematic methods for determining how heterogeneous data sources should be selected, combined, and justified within deployed AI systems.

This gap reflects a mismatch between regulatory expectations and prevailing technical approaches to explainability. Individuals affected by automated decisions rarely seek explanations of internal model mechanics. Instead, they ask which data were used, in what priority, and under whose governance. Existing explainability methods, however, focus primarily on model internals or feature-level attribution \citep{Strumbelj2014,Lundberg2017,Koh2017}. Institutions therefore remain poorly equipped to justify admissible data choices or to translate governance requirements into operational data-selection decisions, as reflected in persistent transparency gaps around training-data documentation and disclosure practices \citep{Wan2025}.

A growing empirical literature shows that many failures attributed to algorithms originate in the composition and governance of training data. Disparate impacts frequently arise from historically patterned datasets \citep{Barocas2016,Mehrabi2021}. Privacy risks increase with data granularity and cross-source linkage \citep{Narayanan2008,Rocher2019}. Robustness failures often reflect narrow or weakly curated training distributions \citep{Recht2019,Taori2020}. Organizational studies further demonstrate how insufficient attention to data practices propagates downstream harms throughout AI development and deployment pipelines \citep{Sambasivan2021}. Taken together, these findings indicate that responsible AI deployment requires structured governance not only of models, but also of the data mixtures that shape model behavior.

To address this problem, we introduce the \emph{Smart Data Portfolio} (SDP) framework, which treats data categories as productive but risk-bearing information assets whose allocation can be governed, audited, and explained. SDPs formalize input governance as an information-risk trade-off defined at the portfolio level. The framework introduces two measurable quantities, Informational Return and Governance-Adjusted Risk, whose interaction generates a Governance-Efficient Frontier describing admissible data mixtures under regulatory constraints. Rather than explaining decisions through internal model mechanics, institutions justify data use by disclosing governed portfolios of authorized inputs. These portfolios are operationalized through Data Portfolio Statements, Data Portfolio Cards, and Consumer Portfolio Reports.

The SDP framework draws on supervisory principles long established in financial regulation, where institutions operate within regulator-defined risk envelopes while retaining discretion over internal models and strategies. Analogously, regulators specify admissible data categories, governance weight bands, and policy risk caps, while institutions optimize data allocations within these boundaries. This separation between regulatory constraint-setting and institutional optimization provides a model-agnostic mechanism for translating governance requirements into operational data practices compatible with existing compliance and risk-management workflows.

The paper makes four contributions. (1) We develop Smart Data Portfolios as an explanation layer for AI systems, enabling decisions to be justified through governed input portfolios rather than model internals. (2) We formalize informational return, governance-adjusted risk, the governance-efficient frontier, and the policy risk cap, establishing a quantitative foundation for input governance. (3) We propose a governance architecture supported by regulatory tools and reporting artifacts that enable transparent and auditable deployment. (4) We illustrate the framework in the telecommunications sector, where heterogeneous applications require distinct governed portfolios within a shared regulatory structure.

The remainder of the paper is organized as follows. Section~\ref{sec:literature} reviews related research. Section~\ref{sec:definitions} defines the SDP framework and its principles. Section~\ref{sec:operationalization} develops the quantitative framework and derives the governance-efficient frontier. Section~\ref{sec:governance} introduces the reporting stack that operationalizes SDPs as an explanation layer. Section~\ref{sec:telecom} presents the sectoral illustration. Section~\ref{sec:conclusion} concludes with implications for AI governance.

\section{Background and Related Literature}
\label{sec:literature}

Research relevant to AI governance has developed across technical, economic, and policy domains, often addressing fairness, privacy, robustness, explainability, and informational efficiency as largely separate problems. These literatures have generated substantial insights into the risks and performance characteristics of machine learning systems, yet they typically evaluate models or datasets in isolation rather than the governed allocation of heterogeneous data inputs. As contemporary regulatory frameworks increasingly require institutions to justify how training data are selected and used, the absence of an integrated approach to managing data composition has become more apparent. This section reviews strands of research that collectively motivate treating training data as governed portfolios of information assets.

\subsection{Transparency and explainability}

An early response to the opacity of machine learning systems focused on explainability at the level of individual predictions. \citet{Strumbelj2014} decompose model outputs into feature-level contributions. \citet{Lundberg2017} provide a unified framework for explaining outputs based on additive feature attributions, while \citet{Koh2017} develop influence functions that trace model behavior back to particular portions of the training set.

Technical approaches to documentation and reporting have complemented these explainability efforts. \citet{Mitchell2019} introduce Model Cards as structured summaries of model behavior, including information about training data, evaluation conditions, and subgroup performance. In parallel, dataset documentation frameworks such as Datasheets for Datasets \citep{Gebru2021} record provenance, composition, collection protocols, and intended uses; \citet{Paullada2021} survey this literature and highlight its role in improving transparency at the level of individual datasets. Taken together, these contributions improve transparency regarding model behavior and individual datasets, while leaving largely unresolved how multiple data sources should be selected, combined, and governed within institutional AI deployments.

\subsection{Fairness and bias}

A large literature examines how machine learning systems can produce unequal or discriminatory outcomes. \citet{Barocas2016} analyze how data collection, feature construction, and labeling choices can embed historical inequities into ostensibly neutral models, leading to disparate impacts even in the absence of explicit use of protected attributes. \citet{Hardt2016} formalize a notion of equality of opportunity in supervised learning and show that standard training objectives can generate systematic differences in error rates across groups, which motivates explicit fairness constraints and post-processing adjustments. \citet{Mehrabi2021} provide a comprehensive survey of bias and fairness in machine learning, cataloging sources of harm that arise from coverage gaps, skewed labels, feedback effects, and mis-specified objectives.

Sectoral studies reinforce these concerns as well. In health care, for example, \citet{Chen2021} review ethical machine learning practices and document how imbalanced or unrepresentative training data can translate into clinically significant disparities in performance across patient groups.

These contributions converge on the view that fairness outcomes depend fundamentally on the composition and coverage of training data, suggesting that governance interventions must operate not only at the level of model outputs but also at the level of data mixtures themselves.

\subsection{Privacy and data protection}

Privacy and data protection have long been central concerns in data-intensive systems. Early work on microdata anonymization showed that combinations of quasi-identifiers can sharply increase the risk of reidentification even when direct identifiers are removed \citep{Samarati2001}. Subsequent analyses of large-scale data practices demonstrate that linkage across heterogeneous sources creates cumulative disclosure risk, often in ways that are difficult to anticipate \citep{Narayanan2008,Rocher2019}. These findings underscore that privacy loss is primarily a function of data structure, granularity, and cross-system interoperability rather than of model design alone.

A complementary line of work develops formal guarantees for limiting individual information leakage. Differential privacy provides mechanisms that bound the inference risk associated with queries or model training and quantifies how privacy loss compounds under composition \citep{Dwork2014,Abowd2018,Wood2020}. This literature makes explicit that privacy exposure depends on the magnitude, sensitivity, and reuse of the underlying data, and that even well-designed models can violate privacy when trained on highly identifying inputs.

Taken together, this literature indicates that privacy risk arises from cumulative decisions about data selection, linkage, and reuse, implying that privacy protection must often be implemented through constraints on how different data categories are combined within training systems.

\subsection{Robustness and data quality}

Robustness to changes in data distributions has been the focus of another active line of work. \citet{Recht2019} construct a test set for ImageNet and show that image classifiers that perform well on the original benchmark can suffer notable accuracy declines when evaluated on their new but closely related dataset. \citet{Taori2020} extend this analysis to a broader range of natural distribution shifts and document systematic performance degradation across models and training regimes. These results suggest that models can be fragile to relatively modest changes in the data generating process, especially when they have been optimized on a narrow distribution of training examples.

In parallel, research on data quality has mapped out common sources of degradation in machine learning procedures. \citet{Jakubik2024} review the state of data quality for machine learning and identify issues such as label noise, incomplete or biased coverage, missing documentation, and the inclusion of manipulated or low-veracity sources. They argue that many observed failures of deployed systems can be traced back to shortcomings in data collection and curation rather than to model architecture. These findings reinforce the view that robustness and reliability are properties of the joint data-model system, highlighting the governance importance of managing training-data composition under evolving deployment conditions.

\subsection{Value of information and efficiency}

A further strand of research considers the value and efficient use of information. Early work in decision analysis, notably \citet{Howard1966}, develops information value theory and shows that the expected benefit of additional information can be quantified in terms of improved decision quality. \citet{Hubbard2014} translates these ideas into practical frameworks for measuring the value of intangibles in organizational settings, emphasizing that many institutions collect more information than they can effectively use.

In finance, \citet{Markowitz1952} introduce the idea that portfolio choice can be framed as a balance between expected return and risk, establishing diversification and risk constraints as central features of decision under uncertainty. Subsequent research develops alternative risk measures that capture tail exposures more effectively, most notably Conditional Value-at-Risk \citep{Rockafellar2000}. Related work on coherent risk measures \citep{Artzner1999} provides axiomatic foundations for evaluating optimal portfolio allocation. \citet{Elton2014} survey the broader development of modern portfolio theory where regulated constraints and model-based estimates jointly determine feasible allocation sets. 

In the context of machine learning, classic arguments emphasize that performance gains often arise from data scale, coverage, and problem formulation rather than from incremental algorithmic refinements \citep{Halevy2009}. \citet{Ghorbani2019} assign values to individual data points or subsets based on their marginal contribution to model performance, building on the Shapley value concept from cooperative game theory \citep{Shapley1953}. Subsequent work extends these ideas to federated learning, data marketplaces, and scalable privacy-aware settings under heterogeneous data \citep{Wang2020,Sun2023,Tian2022,Wang2023}. Collectively, this line of research formalizes information as an asset with heterogeneous and measurable contributions, providing a conceptual foundation for treating training data as allocatable resources subject to risk-return trade-offs.

\subsection{Governance frameworks and emerging principles}

Policy frameworks have begun to synthesize these research strands into articulated governance aims. The OECD Principles on Artificial Intelligence \citep{OECD2019}, the NIST AI Risk Management Framework \citep{NIST2023}, the EU Artificial Intelligence Act \citep{EUAIAct2024}, and the U.S. Blueprint for an AI Bill of Rights \citep{OSTP2022} all emphasize the need for transparency and documentation of AI systems, fairness and non-discrimination in outcomes, privacy and data protection, robustness and security, and proportional use of data and computational resources. They require, to varying degrees, that high-risk AI systems document training data and model behavior, assess subgroup performance, safeguard privacy, and demonstrate robustness under realistic operating conditions.

Despite these advances, existing governance frameworks largely articulate normative objectives rather than operational mechanisms capable of sustaining them. They specify what trustworthy AI systems should achieve, but provide limited guidance on how institutions should organize, constrain, and justify the data choices underlying automated decisions. As a result, training data are frequently assembled through inherited workflows and ad hoc judgments about value and risk, with limited capacity for consistent external scrutiny. This operational gap motivates the Smart Data Portfolio framework developed in this study.

\section{The Smart Data Portfolio Framework}
\label{sec:definitions}

Building on the preceding discussion, the Smart Data Portfolio framework formalizes input governance as a structured allocation problem over regulated data categories. We define a data category as a standardized class of inputs whose internal composition may change over time, but whose use is governed, constrained, and reported as a single entity. Categories can be specified by regulators at an intermediate level of detail, capturing shared provenance, governance-risk characteristics, and semantic scope, rather than raw records or model-specific features. This level of abstraction enables data inputs to be governed in a way that is stable, comparable, and auditable across sectors and institutions.

An SDP represents a constrained allocation of such data categories for model training. Formally, it specifies a vector of portfolio weights that determine how much each category contributes to the overall training mixture. Changing these weights alters two fundamental quantities: informational return and governance-adjusted risk.

Informational return measures the empirically validated performance a model can extract from a given mixture of inputs. The operationalization of informational return is domain-dependent and defined through task-appropriate performance metrics and regulator-approved validation protocols. Governance-adjusted risk, on the other hand, captures the regulatory and institutional exposure generated by a given data portfolio. Unlike model performance, which reflects predictive accuracy, governance risk captures the expected burden associated with deploying a system in real-world, regulated environments. These burdens arise from systematic disparities across affected groups, weaknesses in data provenance and documentation, and instability under foreseeable changes in operating conditions.

Importantly, governance-adjusted risk is a continuous quantity rather than a binary concept. Even strong data-governance hygiene cannot eliminate governance risk. Documentation, lineage tracking, consent management, and minimization requirements ensure procedural compliance but do not equalize the fairness, privacy, provenance, or robustness properties of different data categories. That is why informational return and governance-adjusted risk cannot be collapsed into a single objective or treated as interchangeable. They must be evaluated jointly at the SDP level.

SDPs with different allocations of data categories therefore generate distinct governance-performance trade-offs. As a result, some portfolios are strictly superior, in that no alternative provides higher informational return at the same level of governance-adjusted risk. The set of such portfolios for each level of risk defines the Governance-Efficient Frontier. Institutions typically select portfolios on this frontier to maximize informational return subject to the governance constraints, while regulators control the aggregate risk level.

\begin{figure}[t]
	\centering
	\begin{tikzpicture}[x=12.5cm,y=7cm]
		
		\pgfmathsetmacro{\a}{0.08}
		\pgfmathsetmacro{\b}{0.80}
		\pgfmathsetmacro{\c}{0.70}
		\pgfmathsetmacro{\shift}{0.30}
		\pgfmathsetmacro{\xR}{0.00}
		\pgfmathsetmacro{\yR}{0.18 + \shift}
		
		\pgfmathsetmacro{\xT}{sqrt((\yR - (\a+\shift))/\c)}
		\pgfmathsetmacro{\yT}{\a + \b*\xT - \c*\xT*\xT + \shift}
		
		\draw[line width=0.55pt] (0,0) -- (1.02,0)
		node[below left,yshift=-3.2ex] {Governance-Adjusted Risk};
		\draw[line width=0.55pt] (0,0) -- (0,1.02)
		node[midway,rotate=90,anchor=center,yshift=33pt]
		{Informational Return};
		
		\foreach \x in {0.1,0.2,...,1.0}
		\draw[very thin,gray!25] (\x,0) -- (\x,1.02);
		\foreach \y in {0.1,0.2,...,1.0}
		\draw[very thin,gray!25] (0,\y) -- (1.02,\y);
		
		\foreach \x in {0,0.2,...,1.0}{
			\draw (\x,0) -- ++(0,-0.015) node[below] {\small \pgfmathprintnumber{\x}};
		}
		\foreach \y in {0,0.2,...,1.0}{
			\draw (0,\y) -- ++(-0.015,0) node[left] {\small \pgfmathprintnumber{\y}};
		}
		
		\draw[line width=1.35pt, color=sdpTeal, domain=0.05:1.0, smooth, variable=\x]
		plot ({\x},{\a + \b*\x - \c*\x*\x + \shift});
		
		\node[color=sdpTeal, anchor=north]
		at (0.80,{ \a + \b*0.55 - \c*0.55*0.55 + \shift - 0.15 })
		{\small \textbf{Governance-Efficient Frontier}};
		
		\pgfmathsetmacro{\xCap}{\xT}
		\draw[dashed, line width=1.35pt, color=sdpBlue] (\xCap,0) -- (\xCap,1.02);
		
		\node[color=sdpBlue, anchor=south, rotate=90]
		at (\xCap+0.005,0.20) {\small Policy Risk Cap};
		
		\pgfmathsetmacro{\xA}{0.13}
		\pgfmathsetmacro{\xB}{0.22}
		\pgfmathsetmacro{\xC}{0.66}
		\pgfmathsetmacro{\xD}{0.77}
		
		\pgfmathsetmacro{\yA}{\a + \b*\xA - \c*\xA*\xA + \shift}
		\pgfmathsetmacro{\yB}{\a + \b*\xB - \c*\xB*\xB + \shift}
		\pgfmathsetmacro{\yC}{\a + \b*\xC - \c*\xC*\xC + \shift}
		\pgfmathsetmacro{\yD}{\a + \b*\xD - \c*\xD*\xD + \shift}
		
		\filldraw[fill=sdpTeal, draw=sdpTeal] (\xA,\yA) circle (2.5pt);
		\node[color=sdpTeal, anchor=south] at (\xA,\yA){\small A};
		
		\filldraw[fill=sdpTeal, draw=sdpTeal] (\xB,\yB) circle (2.5pt);
		\node[color=sdpTeal, anchor=south] at (\xB,\yB){\small B};
		
		\filldraw[fill=white, draw=sdpBlue, line width=0.75pt] (\xC,\yC) circle (2.5pt);
		\node[color=sdpBlue, anchor=south] at (\xC,\yC){\small C};
		
		\filldraw[fill=white, draw=sdpBlue, line width=0.75pt] (\xD,\yD) circle (2.5pt);
		\node[color=sdpBlue, anchor=south] at (\xD,\yD){\small D};
		
		\fill[sdpOrange] (\xT,\yT) circle (2.8pt);
		\draw[sdpOrange, line width=1.0pt] (\xT,\yT) circle (3.3pt);
		
		\node[color=sdpOrange, anchor=south, align=center]
		at (\xT,\yT+0.02) {\footnotesize Governance-optimal portfolio};
		
		\foreach \P in {(0.11,0.02+\shift),(0.17,0.07+\shift),(0.24,0.04+\shift)}{
			\filldraw[fill=white, draw=sdpGray, line width=0.75pt] \P circle (2.5pt);
		}
		\node[color=sdpGray, anchor=north] at (0.17,0.00+\shift)
		{\small inferior SDPs};
		
		\node[color=sdpBlue, anchor=north east] at (\xD,\yD) {\small infeasible};
		
	\end{tikzpicture}
	\caption{Illustrative governance-efficient frontier under a policy-defined risk cap. The efficient frontier (teal) traces the highest informational return attainable at each governance-risk level. The policy risk cap (blue dashed line) defines the admissible region. Filled points denote admissible SDPs; hollow points denote dominated or policy-infeasible SDPs. The governance-optimal portfolio (orange) is the highest-return point on the frontier within the risk cap.}
	\label{fig:frontier}
\end{figure}
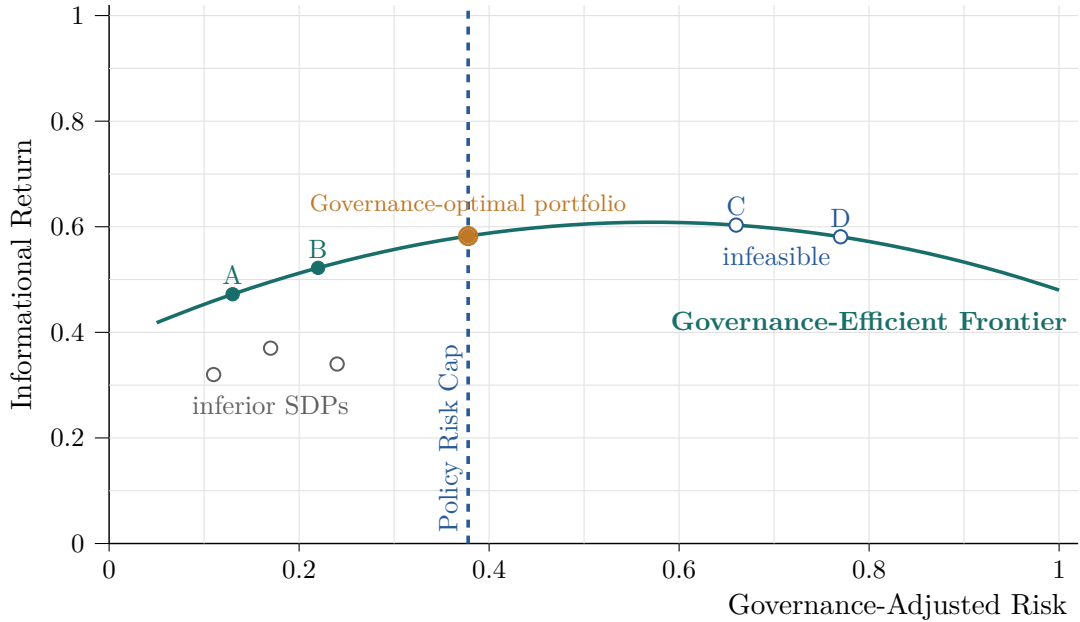

Figure~\ref{fig:frontier} illustrates SDP optimization under a policy-defined risk cap. Each point on the plane corresponds to an SDP, that is, a feasible mixture of data categories. The solid curve represents the Governance-Efficient Frontier, tracing the upper envelope of attainable informational return for each level of governance-adjusted risk. Hollow markers below indicate inferior SDPs that achieve lower informational return than frontier portfolios at comparable risk. The vertical line represents the Policy Risk Cap determined by regulation. Points A and B lie within the admissible region to the left of the Risk Cap, while points C and D lie to the right of the cap and are therefore infeasible. The governance-optimal portfolio is the frontier point that maximizes informational return while remaining within the allowed risk level.

In addition to the policy risk cap discussed above, regulators can shape the design space through Admissible Data Categories as well as Governance Weight Bands. Admissible categories determine which classes of data may be used at all, while weight bands impose upper or lower bounds on their contribution to training mixtures. Together with the policy risk cap, these instruments constrain input composition without prescribing model architectures, training procedures, or optimization methods.

The portfolio-based perspective offers several conceptual advantages. First, it creates measurable governance objects where existing approaches rely largely on qualitative judgment: informational return and governance-adjusted risk become explicit quantities that can be compared across systems and over time. Second, it separates technical performance from governance burden, giving institutions and regulators a shared vocabulary for assessing how data choices shape system behavior. Third, because the logic operates at the level of data categories rather than model architectures, it provides a model-agnostic and sector-neutral input layer for governance that applies across different domains. Fourth, explainability becomes an input-level property, enabling institutions to justify and defend outcomes by referencing approved, risk-bounded data portfolios rather than opaque internal mechanics. Finally, by linking regulatory tools such as risk caps and weight bands directly to portfolio structure, SDPs allow firms and regulators to reason transparently about feasible alternatives when certain data categories must be constrained or excluded.

These ideas are informed by supervisory practices long established in financial regulation, where regulators define admissible risk envelopes while institutions retain discretion over internal methods. The SDP framework adapts this supervisory logic to AI inputs, formalizing how governance parameters shape feasible data allocations without constraining model choice or innovation.

\section{Operationalization}
\label{sec:operationalization}

This section specifies the quantities and procedures needed to implement SDPs as an auditable input-governance layer, including portfolio-level performance and governance-risk measures, the resulting governance-efficient frontier, and the regulatory constraints that define admissible data allocations.

\subsection{Informational Return}

Let \( \mathcal{D}=\{D_1,\ldots,D_n\} \) denote the available data categories, and let \( w=(w_1,\ldots,w_n) \) be a nonnegative weight vector with \( \sum_i w_i=1 \). 

Operationally, the portfolio weight \( w_i \) corresponds to the logged sampling share allocated to data category \( i \) within the training pipeline. Depending on the application, this share may be implemented through record counts, token proportions, minibatch sampling frequencies, or allocated training-compute budgets under the institution’s data-loading policy. Recording these shares renders portfolio composition observable, reproducible, and auditable for supervisory review.

Training a model from class \( \mathcal{M} \) on the mixture induced by \( w \) yields a task-specific performance metric \( M(w;\mathcal{M}) \), evaluated under a governance-approved validation procedure. Depending on the application, this may correspond to AUC, cross-entropy loss, RMSE, calibration error, or other sector-appropriate performance indicators.

We define the informational return of the portfolio as
\begin{equation}
	\mu_p(w) = M(w;\mathcal{M}),
\end{equation}
which captures the empirically validated performance that the institution’s modeling pipeline extracts from the selected mixture of data categories. Informational return is intentionally defined relative to the modeling class \( \mathcal{M} \) as it is not an intrinsic property of the data, but reflects how effectively the institution’s chosen models and training procedures convert inputs into predictive performance.

In practice, institutions estimate \( \mu_p(w) \) using standard validation protocols already embedded in their ML pipelines, such as cross-validation, temporal splits, or other validation procedures accepted within the applicable governance framework. The SDP framework does not prescribe a particular metric or estimation method, requiring only that informational return be computed consistently and documented in a form suitable for regulatory review.

This treatment mirrors modern portfolio theory, where the efficient frontier is defined relative to an institution’s return model and estimation choices rather than as an absolute property of the assets themselves \citep{Markowitz1952}. Accordingly, different institutions may trace different governance-efficient frontiers even under identical regulatory constraints, reflecting heterogeneous modeling strategies rather than governance failures.

\subsection{Governance-Adjusted Risk}

Governance-adjusted risk is a composite measure that captures expected governance burdens. These burdens typically arise from fairness dispersion, documentation and provenance defects, or operational instability. Each of these dimensions is explicitly recognized in contemporary AI governance regimes, including the EU Artificial Intelligence Act \citep{EUAIAct2024}, the OECD AI Principles \citep{OECD2019}, and the NIST AI Risk Management Framework \citep{NIST2023}. To operationalize these concerns at the input level, we decompose governance-adjusted risk into three components:

\begin{enumerate}
	\item \textbf{Fairness Dispersion} (\( F(w;\mathcal{M}) \)). Variation in model performance across relevant subgroups, measured through gaps in error rates, acceptance rates, or domain specific fairness indicators \citep{Barocas2016,Hardt2016}. As with informational return, this is evaluated on the trained model.
	
	\item \textbf{Provenance and Quality Defect Score} (\( P(w) \)). Documentation, provenance, and privacy-governance defects such as missing consent metadata, uncertain licensing, high reidentification exposure due to linkability or granularity, or high label noise \citep{Gebru2021,Narayanan2008,Rocher2019}.

	\item \textbf{Robustness Volatility} (\( R(w;\mathcal{M}) \)). Instability under benign distribution shifts, estimated from out-of-time or out-of-subset tests \citep{Recht2019,Taori2020}. One operational definition is
	\begin{equation}
		R(w;\mathcal{M}) =
		\max_{s\in\mathcal{S}}
		\left|M_{\mathrm{val}}(w;\mathcal{M}) - M_{\mathrm{shift}\;s}(w;\mathcal{M})\right|,
	\end{equation}
	where $\mathcal{S}$ is a regulator-specified set of benign shifts such as temporal drift, mild covariate corruption, or geographic reweighting. This captures operational fragility that can generate governance harm.
\end{enumerate}

For comparability across institutions and to eliminate scale ambiguity, each component is mapped to a common unit interval using regulator-defined scoring rules and transformations. As a result, \( F(w;\mathcal{M}), P(w), R(w;\mathcal{M}) \in [0,1] \), and the relative magnitudes of the policy parameters \( (\alpha,\beta,\gamma) \) reflect regulatory priorities rather than arbitrary metric scales.

While nonlinear interactions are possible, governance-adjusted risk is defined linearly for regulatory interpretability as follows:
\begin{equation}
	\sigma_p(w)=\alpha F(w;\mathcal{M})
	+\beta P(w)
	+\gamma R(w;\mathcal{M}),
\end{equation}

In practice, regulators calibrate the parameters \( (\alpha,\beta,\gamma) \) by mapping each component to its expected institutional cost under applicable law and sectoral penalties. The schedule need not be universal; only stability within a regulatory cycle is required. Continuity of each component implies that the attainable region in the information-risk plane is compact, ensuring that the efficient frontier exists.

The composite quantity $\sigma_p(w)$ is an average-case measure. In high-stakes settings, regulators may instead require tail-sensitive control of rare but severe governance failures, in which case $\sigma_p(w)$ can be replaced by a Conditional Value-at-Risk (CVaR) analogue \citep{Rockafellar2000}.

Let $L(w)=\alpha F(w;\mathcal{M})+\beta P(w)+\gamma R(w;\mathcal{M})$ denote the composite governance loss and let $\operatorname{ES}_\eta$ denote Expected Shortfall at tail probability $\eta\in(0,1)$. The tail-sensitive governance risk is defined as:
\begin{equation}
	\sigma^{\text{CVaR}}_p(w)
	= \operatorname{ES}_\eta\!\bigl[L(w)\bigr]
	= \frac{1}{\eta} \int_{0}^{\eta}\operatorname{VaR}_u\!\bigl[L(w)\bigr]\,du.
\end{equation}
CVaR admits standard convex reformulations, enabling practical implementation in constrained optimization settings \citep{Rockafellar2000}.

\subsection{Governance-Efficient Frontier}

Each portfolio \( w \) generates a point \( \bigl(\mu_p(w),\sigma_p(w)\bigr) \) in the information-risk plane. Varying the weight vector over the simplex defines the attainable region. The governance-efficient frontier is the upper envelope of attainable informational return at each level of governance risk.

In practice, institutions construct the frontier empirically. Starting from a regulator-approved feasible region \( \mathcal{W} \subseteq \Delta^{n-1} \), the procedure can be summarized as follows: 
\begin{enumerate}
	\item Sample candidate portfolios \( w^{(k)} \in \mathcal{W} \) using grid search, Dirichlet sampling, or Bayesian exploration.
	\item For each candidate, estimate the validation metric and governance components by applying the institution’s standard training and evaluation pipeline to the induced mixture.
	\item Compute the portfolio-level quantities
	\begin{equation}
		\mu_p^{(k)} = M(w^{(k)};\mathcal{M}), \qquad
		\sigma_p^{(k)} = \alpha F(w^{(k)};\mathcal{M})
		+\beta P(w^{(k)})+\gamma R(w^{(k)};\mathcal{M}).
	\end{equation}
	\item Discard dominated points and fit the upper envelope of \( \{(\sigma_p^{(k)},\mu_p^{(k)})\} \) by concave hull or smooth
	interpolation to obtain the governance-efficient frontier.
\end{enumerate}

Because informational return and governance-adjusted risk are not statistically dual quantities, the shape of the frontier in \( (\sigma_p(w),\mu_p(w)) \) space is empirical. Evidence summarized in Section~\ref{sec:literature} suggests that, in many regulated consumer-facing settings, the frontier may be approximately concave or inverse-U shaped: marginal informational gains saturate once core categories are included \citep{Recht2019,Taori2020}, while fairness, privacy, and provenance risks typically rise with additional behavioral or weakly curated inputs \citep{Mehrabi2021,Sambasivan2021}. A detailed empirical analysis of frontier shapes and associated confidence intervals is left for future work.

\subsection{Portfolio optimization and regulation}

Once the governance-efficient frontier is established, portfolio optimization proceeds by selecting the data allocation that maximizes validated performance subject to regulator-defined governance constraints. The central regulatory instrument is the \emph{policy risk cap}, which sets an upper bound on aggregate governance exposure while preserving institutional discretion over model architectures, training procedures, and validation strategies. Operationally, the policy risk cap functions as a compliance boundary that determines which data mixtures are admissible for deployment, while leaving model choice and internal training procedures to institutions.

For a given modeling class \( \mathcal{M} \) and policy risk cap \( \bar{\sigma} \), the optimization problem can be written as
\begin{equation}
	\max_w \; M(w;\mathcal{M})
	\qquad \text{s.t.} \qquad
	\sigma_p(w)\le\bar{\sigma},\;
	\sum_i w_i=1,\;
	w_i\ge0.
\end{equation}
This formulation mirrors supervisory practice in financial regulation, where institutions optimize within a regulator-defined risk envelope rather than complying with prescriptive design rules.

In practice, optimization is carried out over a constrained feasible region defined jointly by the policy risk cap and additional category-level governance instruments. Because neighboring portfolios correspond to nearby weight vectors, warm starts and incremental fine-tuning substantially reduce computational cost. For high-dimensional or expensive training regimes, Bayesian optimization with Gaussian Process surrogates provides an efficient exploration strategy \citep{Snoek2012,Shahriari2016}.

\subsubsection*{Admissible data categories}

In many applications, sectoral or jurisdictional governance frameworks impose explicit eligibility rules on the types of data that may be used in AI systems. Within the SDP framework, these rules take the form of \emph{Admissible Data Categories}, which define which classes of data are permitted to enter a portfolio at all. Eligibility reflects legal, ethical, and sector-specific norms related to privacy, documentation, consent, and purpose limitation.

Formally, admissibility constraints operate by excluding specific categories or aggregated groups from the portfolio weight vector \( w \). They define the outer boundary of the feasible region prior to any frontier-based optimization, ensuring that subsequent portfolio selection occurs only within regulator-approved informational boundaries. Typical examples include outright prohibitions or hard limits on high-risk data classes:
\begin{equation}
	\sum_{i\in H} w_i = 0
	\qquad \text{or} \qquad
	\sum_{i\in H} w_i \le \theta_H.
\end{equation}
Such eligibility constraints operate independently of model performance. Even if a prohibited category would yield high informational return, it remains excluded from consideration.

\subsubsection*{Governance weight bands}

While admissible data categories specify which data may be used, \emph{Governance Weight Bands} determine how much of each admissible category may contribute to a portfolio. These further translate qualitative governance objectives into quantitative lower or upper bounds on category-level allocations, playing a role analogous to concentration limits or exposure caps in financial regulation.

Let \( L_i \) and \( U_i \) denote regulator-defined bounds for category \( i \). A governance weight band takes the form
\begin{equation}
	L_i \le w_i \le U_i,
\end{equation}
with \( 0 \le L_i \le U_i \le 1 \). Lower bounds can ensure baseline representation of low-risk or high-integrity data sources, while upper bounds can limit reliance on sensitive, weakly documented, or behaviorally intrusive categories.

Because weight bands act directly on the allocation vector \( w \), they shape the feasible region prior to optimization and prevent circumvention through internal regrouping of data sources. They can also be imposed on aggregated category groups. For example, partitioning categories into low-risk registries \( \mathcal{R} \), operational metrics \( \mathcal{O} \), and behavioral traces \( \mathcal{B} \), a regulator may specify:
\begin{equation}
	\sum_{i \in \mathcal{R}} w_i \ge L_{\mathcal{R}}, \qquad
	\sum_{i \in \mathcal{O}} w_i \in [L_{\mathcal{O}}, U_{\mathcal{O}}], \qquad
	\sum_{i \in \mathcal{B}} w_i \le U_{\mathcal{B}}.
\end{equation}
These constraints ensure minimum reliance on stable data sources, balanced use of operational indicators, and capped exposure to high-risk behavioral data.

\subsubsection*{Interaction with the policy risk cap}

Eligibility rules and governance weight bands operate jointly with the policy risk cap. While admissibility and weight bands constrain \emph{where} portfolios may lie in allocation space, the policy risk cap constrains \emph{how much aggregate governance burden} a portfolio may carry. Together, these instruments define a regulated feasible region within which institutions select the governance-optimal portfolio on the efficient frontier.

This separation of roles allows regulators to enforce high-level governance objectives while leaving institutions free to optimize performance within clearly articulated and auditable boundaries.

\subsection{Attribution and interpretability}

Although the frontier uses the aggregate quantities \( (\mu_p(w),\sigma_p(w)) \), institutions may require transparency regarding the contribution of individual data categories. Because \( M(w) \) is generally nonlinear in \( w \), additive decompositions of the form \( \mu_p \approx \sum_i w_i\mu_i \) lack theoretical justification. As a result, a cooperative game-theoretic formulation is required.

Let \( \mathcal{N}=\{1,\ldots,n\} \) be the set of categories and let \( M(S) \) denote performance when training only on categories \( S\subseteq\mathcal{N} \). The Shapley value for category \( i \) \citep{Shapley1953,Strumbelj2014,Ghorbani2019} is given as:

\begin{equation}
	\phi_i =
	\sum_{S \subseteq \mathcal{N}\setminus\{i\}}
	\frac{|S|!\,(n-|S|-1)!}{n!}
	\big(M(S\cup\{i\}) - M(S)\big).
\end{equation}

Shapley values provide transparent attribution without relying on model-specific heuristics. Importantly, they do not influence frontier construction, which depends solely on aggregate portfolio-level quantities. They instead serve a justificatory role, supporting explanation and contestation of portfolio choices without altering governance constraints or optimization outcomes.

While exact computation is exponential in \( n \), practical estimators for \( \phi_i \) based on Monte Carlo sampling, influence functions, or gradient based methods are well established \citep{Koh2017}. For portfolios with a moderate number of categories, sampled permutations on the order of a few hundred to a few thousand often yield stable estimates in empirical studies \citep{Jia2019}.

\section{Policy and Implementation}
\label{sec:governance}

The Smart Data Portfolio (SDP) framework is designed as an input-level governance architecture rather than a model-level control mechanism. Its central governance contribution is to make data-selection decisions explicit, measurable, and auditable, while preserving institutional discretion over model choice, validation procedures, and optimization routines. By operating at the level of data allocation, SDPs deliver enforceable constraints without prescribing technical implementations. This approach aligns with regulatory guidance emphasizing accountability for data-driven harms \citep{FTC2021} as well as emerging standards such as ISO/IEC 42001 \citep{ISO42001}. This practical focus is especially important for agentic AI systems, whose adaptive and multi-step decision processes can render output-level explanations unstable over time.

\subsection*{Institutional scope and primary users}

The framework is most immediately actionable for three groups. First, sector regulators and supervisory agencies face the task of translating broad statutory mandates into auditable, operational rules for high-risk AI systems. Second, specialized AI-governance consultancies increasingly require quantitative, defensible artifacts to support compliance, auditing, and advisory services across jurisdictions. Third, in-house risk, compliance, and data-governance teams in data-intensive firms bear direct accountability for data-selection decisions and must justify those choices to regulators, auditors, and senior management.

These actors share three structural characteristics: direct responsibility for data governance outcomes, authority to impose portfolio constraints, and a practical need for model-agnostic tooling. By framing data allocation as a governed portfolio-allocation problem, SDPs provide these stakeholders with a common vocabulary and a credible pathway for implementing transparency, fairness, privacy, and robustness requirements across heterogeneous AI systems.

\subsection*{Supervisory workflow and portfolio iteration}

Implementation under the SDP framework proceeds through an iterative supervisory cycle. Regulators specify governance parameters, including admissible data categories, governance weight bands, and an overall policy risk cap. Institutions then construct candidate data portfolios and evaluate them using the informational and governance-risk metrics defined in Section~\ref{sec:operationalization}. Within these constraints, institutions retain full discretion over model class, training procedures, and validation strategies, but must justify resulting input allocations in a structured and reproducible manner.

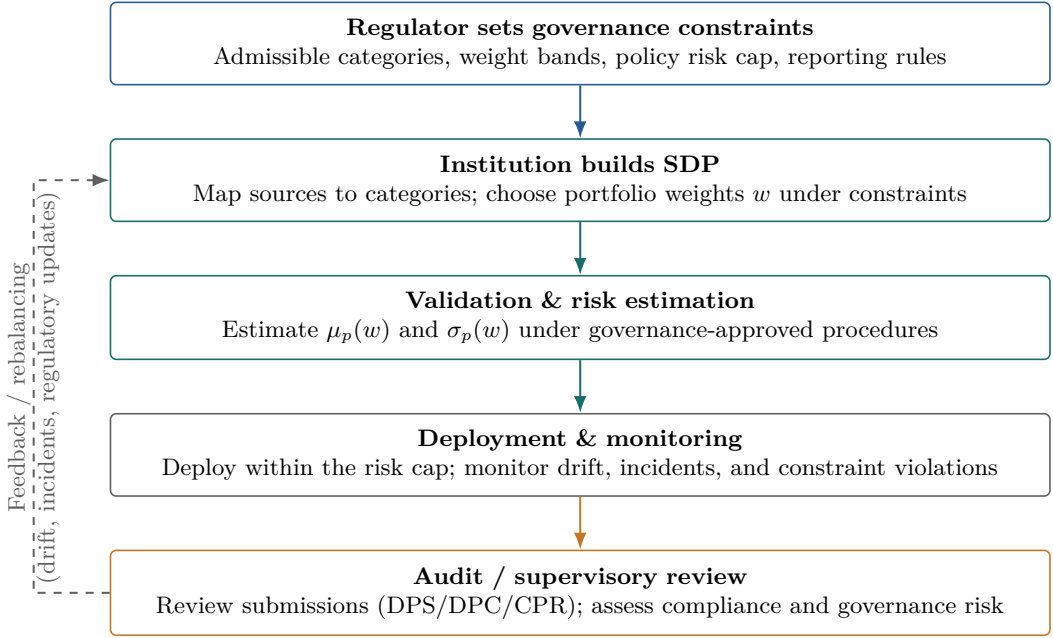
\begin{figure}[t]
	\centering
	\begin{tikzpicture}[
		font=\footnotesize,
		node distance=7mm,
		box/.style={
			draw, rounded corners=2.2pt, line width=0.6pt,
			align=center, inner sep=6pt, text width=0.75\linewidth,
			fill=white
		},
		arrow/.style={-Latex, line width=0.75pt},
		back/.style={-Latex, dashed, line width=0.75pt}
		]
		
		\node[box, draw=sdpBlue] (reg)
		{\textbf{Regulator sets governance constraints}\\
			Admissible categories, weight bands, policy risk cap, reporting rules};
		
		\node[box, draw=sdpTeal, below=of reg] (build)
		{\textbf{Institution builds SDP}\\
			Map sources to categories; choose portfolio weights \(w\) under constraints};
		
		\node[box, draw=sdpTeal, below=of build] (val)
		{\textbf{Validation \& risk estimation}\\
			Estimate \( \mu_p(w) \) and \( \sigma_p(w) \) under governance-approved procedures};
		
		\node[box, draw=sdpGray, below=of val] (deploy)
		{\textbf{Deployment \& monitoring}\\
			Deploy within the risk cap; monitor drift, incidents, and constraint violations};
		
		\node[box, draw=sdpOrange, below=of deploy] (audit)
		{\textbf{Audit / supervisory review}\\
			Review submissions (DPS/DPC/CPR); assess compliance and governance risk};
		
		\draw[arrow, color=sdpBlue] (reg) -- (build);
		\draw[arrow, color=sdpTeal] (build) -- (val);
		\draw[arrow, color=sdpTeal] (val) -- (deploy);
		\draw[arrow, color=sdpOrange] (deploy) -- (audit);
		
		\draw[back, color=sdpGray]
		(audit.west) -- ++(-10mm,0) |- node[
		pos=0.50, left, align=center, text=sdpGray, rotate=90
		]
		{\footnotesize Feedback / rebalancing\\(drift, incidents, regulatory updates)}
		(build.west);
	\end{tikzpicture}
	\caption{Iterative supervisory workflow for Smart Data Portfolios (SDPs), linking regulatory constraint-setting, institutional portfolio construction, deployment monitoring, and review-driven rebalancing.}
	\label{fig:sdp-lifecycle}
\end{figure}

Supervisory oversight occurs through periodic review of submitted portfolios. When constraints are violated or governance thresholds exceeded, regulators may require rebalancing or impose tighter bounds. As data environments evolve through concept drift, the introduction of new data sources, or changes in legal requirements, institutions revisit the portfolio-optimization problem and adjust weights accordingly. This cyclical structure mirrors established practices in financial supervision and model-risk governance, adapted to the specific challenges posed by data-centric AI systems. Figure~\ref{fig:sdp-lifecycle} summarizes the supervisory workflow, emphasizing the iterative loop from constraint-setting to portfolio construction, validation, deployment monitoring, and review-driven rebalancing.

\subsection*{Reporting artifacts and governance objects}

To support this workflow, we propose three complementary reporting artifacts that together form an auditable disclosure trail. These artifacts can be generated using existing workflows, without requiring new technical infrastructure or model instrumentation.

The \emph{Data Portfolio Statement} (DPS) provides a public-facing summary of admissible data categories, applicable governance weight bands, and the principles governing input allocation. The \emph{Data Portfolio Card} (DPC) serves as a regulatory filing that records realized category-level weights together with validated values of informational return \( \mu_p(w) \) and governance-adjusted risk \( \sigma_p(w) \), supported by documentation of data sources, lineage, and validation procedures. The \emph{Consumer Portfolio Report} (CPR) provides consumer-facing explanations of adverse actions or decisions by referencing the approved input portfolio rather than internal model mechanics.

Taken together, these artifacts parallel disclosure regimes used in credit reporting, financial stress testing, and data-protection impact assessments, while being specifically tailored to the governance of AI training data.

\subsection*{Relation to existing documentation tools}

Existing documentation tools, most notably Model Cards and Datasheets \citep{Mitchell2019,Gebru2021}, improve transparency by describing model behavior and individual dataset properties. However, they do not govern how multiple data sources are combined, weighted, or constrained within a training mixture. Smart Data Portfolios operate at this missing allocation layer. While Model Cards summarize model performance and Datasheets document dataset lineage and quality, SDPs provide portfolio-level explanations through standardized reporting artifacts. As AI systems increasingly mediate regulated services, this input-governance layer operates as a primary layer of accountability and explainability in regulated deployments.

Table~\ref{tab:policy-levers} summarizes the core policy instruments within the SDP framework. Each instrument functions at the portfolio level and is designed to complement, rather than replace, existing documentation practices.

\begin{table}[t]
	\centering
	\caption{Policy instruments of the Smart Data Portfolio framework.}
	\label{tab:policy-levers}
	\begin{tabular}{@{}p{5cm}p{9.5cm}@{}}
		\toprule
		\textbf{Instrument} & \textbf{Function within the framework} \\ \midrule
		Admissible data categories &
		Defines which data classes may enter portfolios, reflecting privacy, documentation, and sectoral norms. \\
		Governance weight bands &
		Sets quantitative minimum or maximum allocations for selected categories, analogous to concentration limits or exposure caps in financial regulation. \\
		Policy Risk Cap &
		Establishes an upper bound on governance burden, creating a feasible region for optimization. \\
		Data Portfolio Statement &
		Public summary of category-level weight ranges and data sources. \\
		Data Portfolio Card &
		Filed with regulators; includes validated \( \mu_p \), \( \sigma_p \), and detailed lineage documentation. \\
		Consumer Portfolio Report &
		Provides consumer-facing explanations of decisions based on approved portfolios. \\
		\bottomrule
	\end{tabular}
\end{table}

Beyond these core procedures, the SDP framework can further incorporate a number of supervisory mechanisms informed by long-standing financial-risk governance. These include concentration limits for interconnected data sources, scenario-based stress testing, as well as risk-weighted exposure summaries. Such SDP extensions do not alter the structure of the underlying optimization problem. Appendix~A outlines this broader regulatory design space.

\section{Sectoral Illustration: Telecom as a Multi-Service AI Platform}
\label{sec:telecom}

Telecommunications operators provide a useful illustration of Smart Data Portfolios because they operate multi-service AI platforms built on heterogeneous, high-volume, and governance-sensitive data ecosystems. These environments combine consumer-facing decision systems, personalized services, and engineering optimization tasks within shared data infrastructures subject to stringent privacy and accountability requirements, including constraints on profiling and automated decision-making under European data-protection governance \citep{EDPB2018}. Despite extensive procedural compliance regimes, telecom operators typically receive limited operational guidance on how heterogeneous data sources should be selected, weighted, and constrained within AI training pipelines. The SDP framework clarifies these decisions by expressing input governance as a portfolio-allocation problem.

\subsection{Data categories and governance profiles}

Telecom operators maintain a rich set of data categories that serve different AI applications, including: (i) network KPIs such as latency, jitter, and throughput; (ii) tariff and billing records; (iii) device characteristics; (iv) coarse or granular usage aggregates; (v) location and mobility traces; and (vi) survey-based quality-of-experience (QoE) indicators. These categories differ in informational value and governance risk. Network KPIs and tariff records are structured, high-quality, and low risk. Behavioral telemetry and fine-grained location sequences are informationally rich, but also privacy sensitive, high variance, and susceptible to documentation or provenance defects. 

A portfolio-based governance perspective clarifies these distinctions by mapping each category to contributions in informational return and governance-adjusted risk. Because telecom operators support AI across multiple product lines, a single governance envelope is insufficient. Each AI deployment has a distinct fairness, privacy, documentation, and robustness profile. The Smart Data Portfolio framework assigns each use case its own admissible region and governance-efficient portfolio.

\subsection{Representative AI use cases}

We illustrate this heterogeneity through three representative AI use cases with distinct governance requirements: device finance scoring, personalization, and network QoS prediction. These stylized examples are intended solely to illustrate how governance constraints reshape admissible data allocations under the SDP framework rather than to represent empirical performance claims.

\subsubsection*{High-risk use case: Device finance and credit scoring}

Telecom operators routinely offer handset installment plans and small credit products, placing these models at the intersection of algorithmic fairness requirements and financial regulation \citep{Barocas2016,Hardt2016}. The privacy literature also highlights risks associated with linking behavioral data to financial decisions \citep{Acquisti2016,Wachter2019}. Regulators therefore impose tight governance constraints. In portfolio form, which can be expressed as:
\begin{equation}
	F(w;\mathcal{M}) \le F_{\max}, 
	\qquad 
	P(w) \le P_{\max}, 
	\qquad 
	w_{\text{behavioral}} \le 0.10.
\end{equation}

High-value, low-risk categories typically include payment and billing history, tariff selections, and device attributes. These provide strong predictive signals with limited fairness or privacy exposure. High-risk signals such as granular app telemetry or precise location sequences receive minimal allocation due to fairness and privacy exposure. The governance-efficient frontier identifies the highest attainable informational return consistent with these constraints.

This can be illustrated with a stylized example containing a single high-risk category within an otherwise balanced portfolio. The category-level values shown below represent component scores \(F_i, P_i, R_i\), and their unweighted sum is reported as a Raw risk score; portfolio-level quantities are computed as weighted sums under \(w\):
\begin{equation}
F(w)=\sum_i w_i F_i, \qquad 
P(w)=\sum_i w_i P_i, \qquad
R(w)=\sum_i w_i R_i.
\end{equation}

We consider five data categories (\( i=1,\ldots,5 \)) with weights summing to one. Table~\ref{tab:toy-example1} shows these along with their component risk score \( \sigma_i \).

\begin{table}[t]
	\centering
	\caption{Data categories for portfolio example with a high-risk category}
	\label{tab:toy-example1}
	\begin{tabular}{@{}lcccccc@{}}
		\toprule
		Category & \( w_i \) & \( F_i \) & \( P_i \) & \( R_i \) & Raw risk score & Note \\ \midrule
		Payment/billing history & 0.40 & 0.02 & 0.01 & 0.03 & 0.06 & Low risk \\
		Tariff selections       & 0.25 & 0.01 & 0.01 & 0.02 & 0.04 & Low risk \\
		Device attributes       & 0.15 & 0.02 & 0.01 & 0.02 & 0.05 & Low risk \\
		Coarse usage patterns   & 0.15 & 0.04 & 0.03 & 0.05 & 0.12 & Medium risk \\
		Behavioral traces       & 0.05 & 0.07 & 0.06 & 0.08 & 0.21 & High risk \\
		\bottomrule
	\end{tabular}
\end{table}

Assume policy parameters \( \alpha=\beta=\gamma=1/3 \). Then the governance-adjusted risk of the portfolio is:
\[
\sigma_p(w) = 
\tfrac{1}{3}F(w)+
\tfrac{1}{3}P(w)+
\tfrac{1}{3}R(w).
\]

Aggregating across categories yields an overall risk of approximately \( \sigma_p(w)=0.081 \), which falls below a hypothetical policy risk cap of \( \bar{\sigma}=0.10 \). At the same time, the behavioral category weight constraint \( w_{\text{behavioral}}\le 0.10 \) binds. A neighboring candidate portfolio with \( w_{\text{behavioral}}=0.12 \) would achieve slightly higher informational return but would violate governance constraints and is therefore ineligible. This demonstrates how governance constraints reshape the attainable frontier by excluding portfolios that rely too heavily on sensitive or weakly curated signals.

The Data Portfolio Statement (DPS) summarizes weight bands and public category-level rationales, while the Consumer Portfolio Report (CPR) provides adverse-action explanations consistent with explainability principles.

\subsubsection*{Medium-risk use case: Personalization and recommendation systems}

Many telecom operators support recommendation services through video, music, social-media bundles, and personalized offers. These applications raise significant privacy and provenance concerns. Behavioral telemetry is often sensitive and disproportionately revealing \citep{Acquisti2016}. 

Because contestable explanations are essential in consumer-facing services, particularly where recommendations may affect vulnerable users, a regulator may impose governance weight bands such as:
\begin{equation}
	w_{\text{behavioral}} \le 0.15,
	\qquad 
	w_{\text{QoS}} \ge 0.20,
	\qquad 
	P(w) \le P_{\max}.
\end{equation}

Here, governance risk is dominated by privacy sensitivity and robustness volatility, given evidence of instability in content recommendation and multimodal systems under distribution shift \citep{Recht2019,Taori2020}. The governance-efficient portfolio therefore emphasizes coarse usage aggregates, network KPIs, and minimally sensitive survey or preference data.

The resulting medium-risk portfolio can be illustrated with a stylized example shown in Table~\ref{tab:toy-example2}. As above, the table entries represent component scores \(F_i, P_i, R_i\).

\begin{table}[h!]
	\centering
	\caption{Data categories for portfolio example with a medium-risk category}
	\label{tab:toy-example2}
	\begin{tabular}{@{}lcccccc@{}}
		\toprule
		Category & \( w_i \) & \( F_i \) & \( P_i \) & \( R_i \) & Raw risk score & Note \\ \midrule
		QoS KPIs & 0.25 & 0.01 & 0.01 & 0.03 & 0.05 & Required minimum \\
		Survey preferences & 0.20 & 0.02 & 0.01 & 0.02 & 0.05 & Low risk \\
		Coarse usage aggregates & 0.25 & 0.04 & 0.03 & 0.05 & 0.10 & Medium risk \\
		Behavioral traces & 0.15 & 0.06 & 0.06 & 0.07 & 0.19 & Weight cap binds \\
		Contextual metadata & 0.15 & 0.03 & 0.02 & 0.03 & 0.08 & Medium risk \\
		\bottomrule
	\end{tabular}
\end{table}

Assuming the same policy parameters as above, the aggregate risk is \( \sigma_p(w)\approx 0.076 \), below the hypothetical \( \bar{\sigma}=0.10 \), and the behavioral cap binds at \( w_{\text{behavioral}}=0.15 \). Compared to the device-finance example, the constraints are milder, which can lead to a governance-optimal portfolio at a relatively close point to the unconstrained frontier. The Data Portfolio Card (DPC) records weight bands, tested values of \( \mu_p(w) \) and \( \sigma_p(w) \), and documentation of consent, provenance, and lineage quality.

\subsubsection*{Low-risk use case: Network QoS prediction and hazard models}

Network operators rely on AI to predict congestion, outages, and quality degradation. These models primarily use engineering signals and operational logs, which carry lower governance risks relative to consumer-level data. Because the data are mainly operational and engineering-driven, governance concerns relate mostly to reliability rather than distributional fairness \citep{Recht2019,Jakubik2024}.

In this case, regulators may impose light constraints such as:
\begin{equation}
	w_{\text{network\_KPIs}} \ge 0.40,
	\qquad 
	R(w;\mathcal{M}) \le R_{\max}.
\end{equation}

Because governance risk is low, the governance-efficient portfolio closely approximates the unconstrained efficient frontier. This contrast illustrates the flexibility of SDPs, which can both tighten or relax constraints depending on use-case risk without altering the underlying optimization structure.

\subsection{Comparing Portfolio Implications Across Use Cases}

The three cases discussed above illustrate a natural risk gradient across telecom AI deployments: financial decisions impose the tightest governance bands, personalized services require moderate constraints, and engineering applications operate under lighter oversight. Table~\ref{tab:telecom-summary} summarizes how distinct governance priorities translate into different portfolio structures.

\begin{table}[h!]
	\centering
	\caption{Governance profiles and data portfolio implications across telecom AI use cases.}
	\label{tab:telecom-summary}
	\begin{tabular}{@{}p{3.1cm}p{4.2cm}p{6.5cm}@{}}
		\toprule
		\textbf{Use case} 
		& \textbf{Governance priorities} 
		& \textbf{Portfolio implications} \\ \midrule
		
		Device finance / credit scoring  
		& Fairness, privacy, auditability  
		& Prioritize payment histories, tariffs, device data; minimal behavioral telemetry; tight policy risk cap. \\[0.3em]
		
		Personalization / recommendations  
		& Privacy, child safety, explainability  
		& Limit granular telemetry; emphasize QoS aggregates and coarse usage groups; behavioral weight cap binds. \\[0.3em]
		
		Network QoS prediction  
		& Robustness, provenance, operational reliability  
		& High allocation to engineering logs and KPIs; fairness constraints minimal; near-unconstrained frontier optimum. \\ \bottomrule
	\end{tabular}
\end{table}

As the table shows, SDPs provide a structured governance discipline that links regulatory objectives to operational data selection decisions. They can also improve explainability by bridging the gap between high-level regulatory principles and operational data selection. Although illustrated in telecom, the same governance logic extends to any domain where heterogeneous data categories carry differentiated risk, including finance, insurance, health, logistics, and mobility services.

\section{Conclusion}
\label{sec:conclusion}

Artificial intelligence systems increasingly operate within regulatory environments that place explicit responsibility on institutions to govern the training data underlying automated decision-making. While contemporary frameworks require documentation, representativeness, and risk management of datasets, they provide limited operational guidance for translating these obligations into implementable data-selection practices. The Smart Data Portfolio framework addresses this institutional gap by introducing a model-agnostic approach to governing AI training data through structured allocation and measurable governance risk.

By conceptualizing data categories as risk-bearing informational assets, SDPs make input governance observable, comparable, and auditable. Portfolio-level quantities linking informational performance to governance exposure allow regulators to constrain aggregate risk while preserving institutional flexibility in model design and deployment. In this framework, explainability shifts from post hoc inspection of model internals toward transparent justification of governed data use. Reporting instruments such as Data Portfolio Statements, Data Portfolio Cards, and Consumer Portfolio Reports translate portfolio choices into accountability mechanisms intelligible to regulators, auditors, and affected individuals.

The framework is designed for integration within existing compliance and risk-management environments. Because SDPs operate at the level of data allocation rather than model architecture, they remain compatible with heterogeneous AI systems and evolving technical practices. Sectoral illustrations demonstrate how diverse applications can operate under shared governance principles while maintaining application-specific data portfolios.

More broadly, Smart Data Portfolios suggest that input governance can be treated as a supervisory discipline analogous to risk management in finance. As implementation of the EU Artificial Intelligence Act and related governance frameworks progresses, operational methods for managing training data will become increasingly necessary \citep{EUAIAct2024,NIST2023}. Future research should therefore examine empirical estimation of governance-efficient frontiers, institutional adoption pathways, and auditing procedures under real-world regulatory constraints. By providing a quantitative yet institutionally interpretable framework for governing training data, SDPs contribute toward scalable and transparent accountability in large-scale AI deployment.

\appendix
\section*{Appendix A: Finance-Inspired Extensions of Smart Data Portfolios}
\label{appendix:extensions}

This appendix outlines several supervisory mechanisms that can be integrated into Smart Data Portfolios without altering the underlying optimization structure. Each mechanism has a conceptual analogue in long-standing financial-risk governance and provides regulators with additional levers for shaping data allocation. They are intentionally illustrative rather than exhaustive and are not required for baseline SDP implementation.

\subsection*{A.1 Concentration limits for connected data sources}

Supervisory regimes often restrict concentrated exposure to a single counterparty. An analogous constraint for training data prevents excessive reliance on a single vendor, broker, or provenance chain. For each equivalence class $G_k$ of connected data suppliers,
\begin{equation}
	\sum_{i\in G_k} w_i \;\le\; \theta_k,
\end{equation}
where $\theta_k$ is regulator-set. These linear constraints complement the weight-band structure in the main text and guard against systemic dependence on high-risk intermediaries \citep{BCBS2832014}.

\subsection*{A.2 Scenario-based governance stress testing}

In analogy with supervisory stress testing (e.g.\ DFAST), regulators may evaluate portfolios under a set of hypothetical or historically informed governance scenarios $s\in\mathcal{S}$, such as demographic shifts, sensor failures, or abrupt changes in privacy law:
\begin{equation}
	(\mu^{(s)}_p(w),\sigma^{(s)}_p(w)).
\end{equation}
A stress-based admissibility condition requires $\sigma^{(s)}_p(w)\le\bar{\sigma}^{(s)}$ for all $s\in\mathcal{S}$. This approach exposes latent fragility in ways not captured by average-case risk measures \citep{FedStressScenarios2024}.

\subsection*{A.3 Risk-weighted data exposures (D-RWA)}

Supervisors may summarize portfolio exposure using a single linear index analogous to bank risk-weighted assets (RWA):
\begin{equation}
	\operatorname{D\!-\!RWA}(w)=\sum_i \rho_i\, w_i,
\end{equation}
where $\rho_i$ is a regulator-defined risk weight calibrated to fairness exposure, provenance uncertainty, or privacy sensitivity. A constraint $\operatorname{D\!-\!RWA}(w)\le B$ provides an interpretable, scalable tool for sector-wide monitoring.

\subsection*{A.4 Governance baselines via risk-parity weights}

A natural baseline for benchmarking institutional portfolios is an equal-risk-contribution allocation \citep{Maillard2010}, defined as the solution to
\begin{equation}
	w_i\,\frac{\partial \sigma_p}{\partial w_i}
	= w_j\,\frac{\partial \sigma_p}{\partial w_j},
	\qquad\forall i,j.
\end{equation}
Risk-parity weights provide a regulator-friendly neutral reference against which deviations can be audited or justified.

These extensions demonstrate the compatibility of Smart Data Portfolios with established supervisory traditions while preserving the framework’s generality and transparency.

\bibliographystyle{elsarticle-harv}
\bibliography{yalta-smart-data-portfolio}

\end{document}